\documentclass[aps,prb,twocolumn,showpacs,floatfix,a4paper]{revtex4-2}
\usepackage{graphicx}
\usepackage[intlimits]{amsmath}
\usepackage{amssymb}
\usepackage{color}

\DeclareMathOperator{\rank}{rank}
\usepackage{expl3}
\ExplSyntaxOn
\pdf_version_gset:n{1.7}
\ExplSyntaxOff

\begin{document}

\title{Ubiquitous Flat Bands in Edge-Inflated Random Graphs}

\author{Richard Berkovits}
\affiliation{Department of Physics, Jack and Pearl Resnick Institute, Bar-Ilan University, Ramat-Gan 52900, Israel}

\begin{abstract}
Flat bands are usually associated with periodic lattice geometries, where destructive interference or chiral-symmetry constraints produce macroscopically degenerate eigenstates. Here we show that flat bands arise generically in a broad class of nonperiodic random networks through a simple edge-inflation procedure. Starting from random regular, Voronoi tessellation, Delaunay triangulation, scale-free, small-world, and Erd\"os-R\'enyi seed graphs, each edge is replaced by a finite one-dimensional tight-binding chain, either with fixed length $L$ or with random lengths of mean $\bar L$. Although the original random graphs typically have adjacency matrices of full rank and therefore few or no zero eigenstates, their inflated counterparts exhibit extensive flat bands. The flat-band energies are determined by the eigenstates residing on the chains or, more generally, by the eigenvalues of identical loopless motifs replacing the seed bonds. The corresponding eigenstates are states supported on subsets of the inflated chains formed by destructive interference at the original seed vertices. For zero-energy flat bands, we relate the degeneracy to the graph nullity $\eta(G)$ and the maximum matching number $\nu(G)$. The numerical results support the correspondence $\eta(G)=N-2\nu(G)$, up to small deviations, across the random graph ensembles studied. The fraction of degenerate zero-energy states in inflated random graphs is controlled primarily by the average degree of the seed graph. We further show that the flat bands exhibit characteristic robustness to disorder: disorder on the seed sites and random magnetic flux leave the flat bands essentially unchanged, while bond disorder and disorder on the chain or motif sites broaden them in a predictable manner. These results show that flat bands do not require crystalline order, but can emerge as robust consequences of local graph inflation, motif substitution, and connectivity constraints. 
\end{abstract}

\maketitle

\section{Introduction}
\label{s1}

Flat bands in tight-binding systems can be viewed as a manifestation of an extreme form of spectral degeneracy, where a macroscopic number of eigenstates collapse to the same energy. From this perspective, the key question is not only how such bands arise, but what structural constraints are sufficient to enforce them. In many known examples, flat bands emerge when the connectivity of the lattice imposes linear dependencies among the hopping amplitudes, effectively reducing the rank of the Hamiltonian. This leads to an extensive kernel of states that are decoupled from the dispersive spectrum. 

A particularly transparent realization of this idea is provided by destructive interference, where amplitudes cancel at specific sites, yielding compact localized states (CLS) \cite{Sutherland1986,Lieb1989,Mielke1991,Tasaki1992,Ramachandran2017,Mallick2022,Maimaiti2017,Maimaiti2019,Maimaiti2021,Kim2025}. However, this mechanism can be understood more broadly as a consequence of algebraic constraints on the underlying graph rather than fine-tuned phase relations. The robustness of these flat bands to various types of disorder was also demonstrated \cite{vidal1998,vidal2001,Bodyfelt2014,Leykam2017,Danieli2015,Cadez2021,Kim2023}. As a result, flat bands appear across a wide range of systems, from Hubbard models exhibiting flat-band ferromagnetism \cite{Lieb1989,Mielke1991,Tasaki1992} to photonic, cold-atom, and engineered electronic platforms \cite{Xia2018,Leykam2018}. This broader viewpoint suggests that flat bands are not exceptional features tied to special lattices, but rather generic outcomes of certain structural redundancies in the connectivity of tight-binding networks.

Results on random graphs show that adjacency matrices (which correspond to tight-binding Hamiltonians with vanishing on-site energies) are generically of full rank, with nonzero nullity (i.e., zero eigenvalues) arising only from specific structural features such as isolated vertices or finely tuned local dependencies \cite{Bordenave2011,Huang2021}. In particular, for random regular graphs, where all vertices have the same degree and isolated sites are absent, the probability of having zero eigenvalues vanishes in the large-size limit.

\begin{figure*}
\centering
    \includegraphics[width=0.8\textwidth]{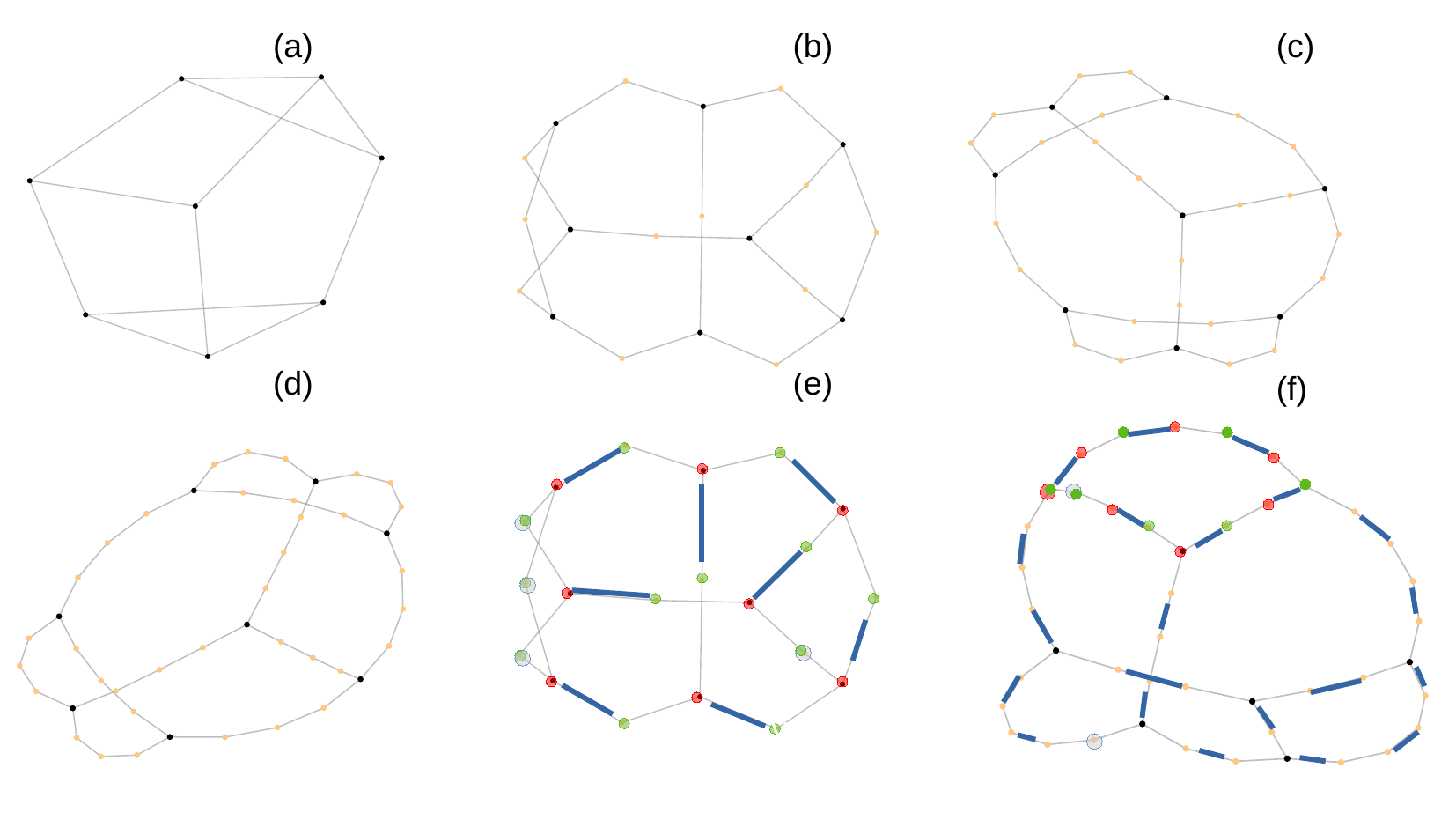}
    \caption{An edge-inflated random regular graph (RRG). (a) The seed RRG, with $N_o=8$ nodes and degree $k=3$. (b) An inflated RRG, in which each edge between the original RRG sites (black circles) is replaced by two edges connected through a newly added site (orange circles). (c,d) The inflated RRG in which each original edge is inflated twice, $L=2$, and three times, $L=3$, respectively. The red and green circles illustrate a bipartition into sublattices A and B, while the thick blue lines denote a matching between pairs of vertices, yielding the maximum possible coverage of the graph. (e) A regularly inflated RRG with $L=1$, which is bipartite, with $|N_B-N_A|=4$, maximum matching $\nu(G)=8$, and $|N_B-N_A| = N - 2\nu(G)$. Unmatched nodes are indicated by a gray halo. (f) A randomly inflated RRG with average inflation length $\bar L=3$. This graph cannot be bipartitioned, as illustrated by the frustrated node on the upper loop. Nevertheless, $\nu(G)=21$ and $N - 2\nu(G) = 2$.}
    \label{fig1}
\end{figure*}

Here we show that inflating the edges of different classes of random graphs, i.e., replacing each edge by a one-dimensional tight-binding chain (see Fig.~\ref{fig1}) or other motifs, gives rise to flat bands containing a macroscopic number of states. The inflation procedure may be applied uniformly to all edges of the seed graph, or sequentially and randomly to selected groups of edges according to a prescribed rule. After inflation, the resulting graph retains the large-scale connectivity of the seed graph and may exhibit a broad distribution of chain lengths. Thus, a simple modification of random graphs, which generically possess very few zero-energy states, can transform them into graphs with an extensive number of such states forming flat bands. As shown below, the number of zero-energy states in a given graph can be accurately accounted for by structural properties of the graph.

\section{Inflation}
\label{s2}

Specifically, we start from a seed random graph $G_o=(V_o,E_o)$, where each site is a node $v_o\in V_o$, the number of original nodes is $N_o=|V_o|$, and each allowed hopping process is represented by an edge $(u_o,v_o)\in E_o$. The number of original edges is denoted by $M_o=|E_o|$.

Edge inflation is a process in which edges of the original graph are rewritten by inserting additional nodes along them. In the elementary step, an edge $(u_j,v_j)$ of the seed graph $G_o$, with $1\le j\le M_o$, is chosen and replaced by two edges, $(u_j,v_p)$ and $(v_p,v_j)$, where $v_p$ is a newly added node.

This procedure may be performed either uniformly or randomly. In the uniform case, $L$ new nodes are inserted along each of the $M_o$ original edges, resulting in $N=N_o+L M_o$ nodes. In the random case, one of the original edges is chosen at random at each step, and this procedure is repeated for $K$ steps. The number of inserted nodes therefore varies from edge to edge, with an average value $\bar L=K/M_o$, so that $N=N_o+\bar L M_o$ on average. Thus, for large $M_o$ and $K$, the probability of obtaining a chain of length $L$ is \cite{Berkovits2026}
\begin{equation}
P(L) \to e^{-\bar L}\,
\frac{\left(\bar L\right)^{L}}{L!},
\qquad L=0,1,2,\dots,
\label{Ldis}
\end{equation}
Here, $L$ denotes the number of inserted nodes in each chain, while $\bar L$ denotes the average number of inserted nodes per chain.

The resulting edge-inflated graph is denoted by $G=(V,E)$, where each site is a vertex $v\in V$ and each edge is a pair $(u,v)\in E$. For uniform inflation, the graph contains $N=N_o+L M_o$ nodes and $(L+1)M_o$ edges.

\section{The spectrum of inflated graphs}
\label{s3}

In the single-particle tight-binding description, the Hilbert space is spanned by the site states $\{\lvert v\rangle\}_{v\in V}$, and the Hamiltonian is
\begin{equation}
H = \sum_{v\in V}\epsilon_v\,|v\rangle\langle v|
+ \sum_{(u,v)\in E}\left(-t_{uv}\,|u\rangle\langle v| - t_{vu}\,|v\rangle\langle u|\right).
\label{H}
\end{equation}
The on-site energies $\epsilon_v$ determine the diagonal terms, while the graph connectivity determines the nonzero off-diagonal matrix elements and the hopping amplitudes $t_{uv}$ determine their values. Unless otherwise stated, $t_{uv}=t$ on all edges and $\epsilon_v=0$, so $H$ reduces to $-t$ times the adjacency matrix of $G$. For the disordered cases considered in Sec.~\ref{s6}, $\epsilon_v$ and $t_{uv}$ may be chosen from a distribution, and the hopping amplitudes may acquire a random phase, $t_{uv}=|t_{uv}| \exp{i\theta_{uv}}$, while $t_{vu}=t^*_{uv}$.

\begin{figure*}[t]
\centering
\includegraphics[width=0.95\textwidth]{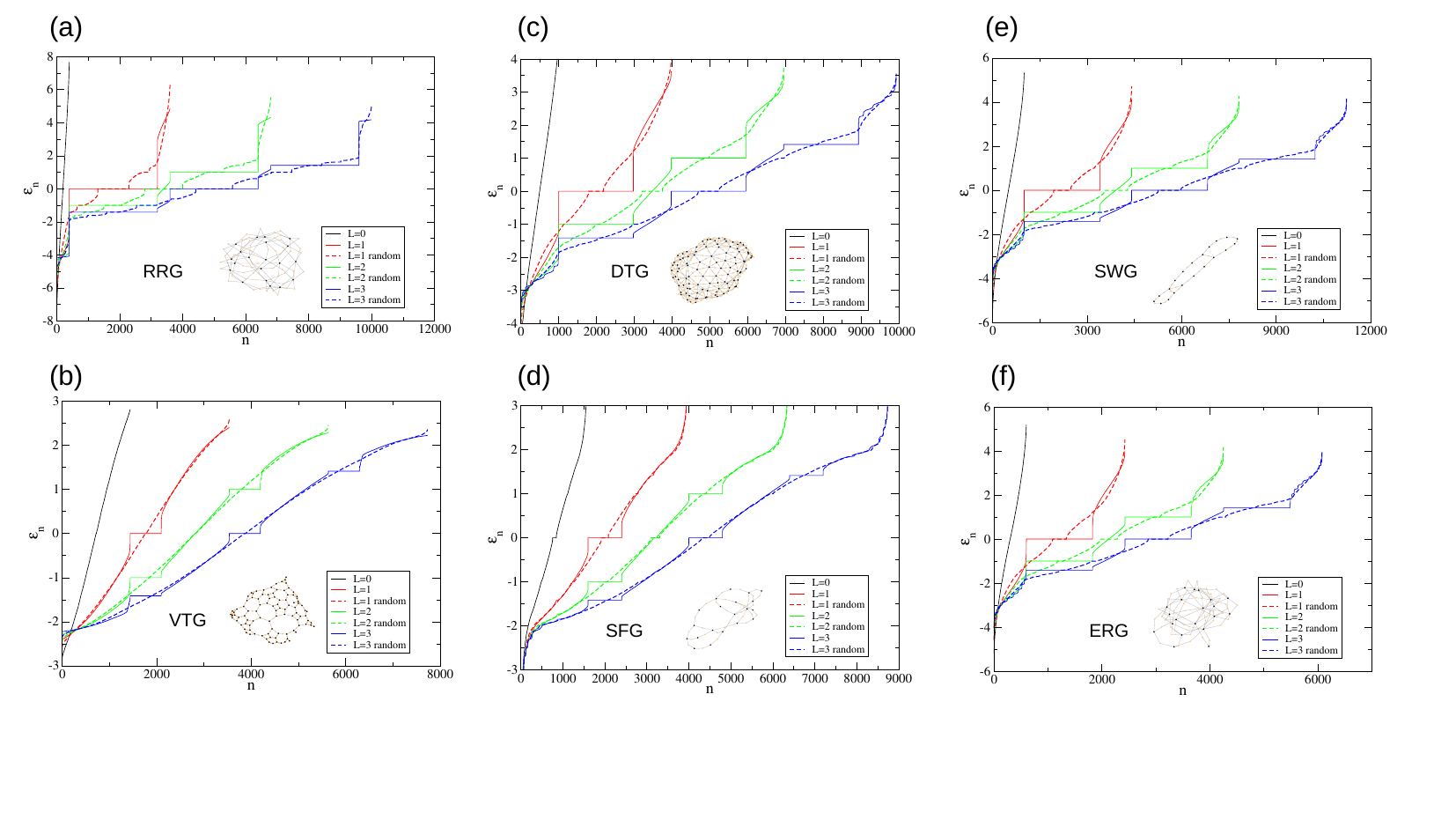}
\caption{The $n$-th energy level $\varepsilon_n$ as a function of the level index $n$ for different types of inflated random graphs. In all cases, an initial seed graph of size $N_o$ is inflated by inserting additional nodes and edges. For uniform inflation, $L$ additional edges and vertices are inserted along each original edge of the graph. For random inflation, the seed edges are inflated randomly so that the average number of additional edges between the original sites is $\bar L$. Shown are: (a) an inflated random regular graph (RRG) with seed size $N_o=400$, and $k=16$; (b) an inflated Voronoi tessellation graph (VTG) with $N_o=1440$; (c) an inflated Delaunay triangulation graph (DTG), which is the dual graph of the corresponding VTG, with $N_o=1000$; (d) an inflated scale-free graph (SFG) with $N_o=1600$ and $\gamma=2.9$; (e) an inflated small-world graph (SWG) with $N_o=1000$ and $p=0.005$; and (f) an inflated Erd\"os-R\'enyi graph (ERG) with $N_o=600$ and $p=0.01$. In each panel, results are shown for both fixed chain lengths $L=1,2,3$ and random chain lengths with average values $\bar L=1,2,3$.}
\label{fig2}
\end{figure*}


To examine the extent to which the spectral properties depend on the topology of the underlying seed network, we consider several representative classes of random graphs that are widely used in statistical physics and network theory \cite{albert2002,Dorogovtsev2002,newman2003}. Fig.~\ref{fig2} shows the ordered energy levels, $\varepsilon_n$, as a function of the level index $n$ for inflated seed graphs drawn from the following ensembles: random regular graphs (RRGs), in which all nodes have the same degree \cite{Wormald1999,McKay1990,bollobas2001,tikhonov2021}; Voronoi tessellation graphs (VTGs), which are spatially embedded random cellular networks generated from a random set of points \cite{okabe2000,aurenhammer1991,Moller1994,Chiu2013,weaire1984,rivier1985,peshkin1991}; Delaunay triangulation graphs (DTGs), which are the dual graphs of the corresponding Voronoi tessellations and provide a standard realization of topological disorder in random lattices \cite{okabe2000,aurenhammer1991,puschmann2015}; scale-free graphs (SFGs), characterized by a power-law degree distribution $P(k)\sim k^{-\gamma}$ \cite{barabasi1999,albert2002,Dorogovtsev2002,newman2003}; small-world graphs (SWGs), which combine strong local clustering with short characteristic path lengths \cite{watts1998,newman2003}; and Erd\"os-R\'enyi graphs (ERGs), the classical ensemble in which edges are placed independently at random \cite{erdos1959,erdos1960,newman2001,bollobas2001,newman2003}.

A random regular graph (RRG) is a homogeneous random network in which each node has exactly the same degree, $k$. Here, for the numerical studies, we chose $k=16$, $N_o=400$, resulting in $M_o=k N_o/2=3200$. Thus, unlike Erd\"os-R\'enyi graphs, RRGs do not have degree fluctuations; their disorder originates solely from the random pairing of edges subject to the fixed-degree constraint. In the large-system limit, such graphs are locally tree-like and therefore provide a useful model for studying spectral statistics, Anderson localization, and high-dimensional connectivity effects \cite{McKay1990,Wormald1999,bollobas2001,tikhonov2021}. The Erd\"os-R\'enyi graph, by contrast, is the classical random graph ensemble in which each pair of nodes is connected independently with a fixed probability $p$, which here was set as $p=0.01$, $N_o=600$, $M_o=p N_o(N_o-1)/2 =1797$ and $\langle k \rangle \sim p N_o = 6$. For the parameters considered here, the binomial degree distribution is well approximated by a Poisson distribution with mean $\langle k\rangle\simeq 6$. For the parameters considered here, the binomial degree distribution is well approximated by a Poisson distribution with mean $\langle k\rangle\simeq 6$, and it serves as a basic reference model for uncorrelated random connectivity \cite{erdos1959,erdos1960,newman2001,newman2003}. In the numerics we accept only realizations of the ERG which are connected.

The Voronoi tessellation graph (VTG) provides a spatially embedded example of a disordered graph. It is generated by first choosing a set of random points in the two-dimensional plane and constructing the corresponding Voronoi tessellation, which partitions space into cells such that all points inside a given cell are closer to its generating point than to any other generator. The graph used here is the one-skeleton of this tessellation, namely the network formed by the vertices and edges of the Voronoi cells. Such graphs naturally describe random cellular structures, including foams, grain boundaries, amorphous materials, and other disordered spatial networks \cite{Moller1994,okabe2000,aurenhammer1991,weaire1984,rivier1985,peshkin1991}. Here $N_o=1440$, and since vertexes in The VTG have a degree of $3$, $M_0 \leq 3 N_o/2 = 2160$. Closely related Delaunay lattices, which are the geometric dual of the Voronoi construction where two generating points are connected by an edge whenever their Voronoi cells share a boundary. Equivalently, the Delaunay triangulation connects nearby points so that the circumcircle of each triangle contains no other generating point. Delaunay lattices have been used in studies of electronic states and Anderson localization on topologically disordered lattices \cite{puschmann2015}. In our simulations we chose $N_o=1000$ and since for the DTG the average degree is $6$ resulting in $M_o\leq 3 N_o = 3000$.

Scale-free graphs (SFGs) are characterized by a broad, power-law degree distribution, $P(k)\sim k^{-\gamma}$, where $\gamma$ is the degree exponent, here chosen as $\gamma=2.9$, $N_o=1600$, $M_o=2400$, minimal degree of $2$ and average degree of $\langle k \rangle = 3$. This distribution leads to strong connectivity inhomogeneity and the emergence of highly connected hub nodes, which can have a significant influence on transport, spectral properties, and localization phenomena \cite{barabasi1999,albert2002,newman2003}. Small-world graphs (SWGs), on the other hand, interpolate between ordered lattices and random graphs. They combine strong local clustering with short characteristic path lengths generated by long-range random links. Initially all sites are connected as a circle, while sites are chosen with probability $p$ and connected by additional edges (here $N_o=1000$, $p=0.005$, $M_o=N_o (1+ p (N_o-3)/2) = 3492$, and $\langle k \rangle = 2 M_o/N_o \sim 2 + p N_o = 7$), SWG provide a useful model for networks that are neither fully regular nor fully random \cite{watts1998,newman2003}. Erd\"os-R\'enyi graphs (ERGs) combine both degree fluctuations (though smaller than in SFGs) and random pairing of edges.

Despite the substantial differences between these graph ensembles, such as fixed degree in RRGs, 2D embedding with no crossing of edges in VTGs and DTGs, power-law degree heterogeneity in SFGs, high clustering and short paths in SWGs, and Poissonian random connectivity in ERGs, their inflated versions exhibit qualitatively similar spectral features. Increasing either the fixed filament length $L$ or the mean random filament length $\bar L$ introduces additional one-dimensional structure into the graph and reorganizes the spectrum accordingly.

For each of the above classes of random graphs, we generate $100$ independent seed graphs. Each realization of $G_o$ is then inflated to produce chains with $L=1,2,3$ nodes in the case of ordered consecutive inflation, and chains with random lengths whose average number of nodes per chain is $\bar L=1,2,3$ in the case of random inflation. For each resulting graph $G$, we exactly diagonalize the corresponding Hamiltonian in Eq.~\ref{H} to obtain its eigenvalues $\varepsilon_n$.

\section{Origin of flat bands}
\label{s4}

A useful way to view the zero-energy flat bands is not to start from global bipartiteness, but from its limitations. For a bipartite graph, the Hamiltonian has chiral symmetry, and a sublattice imbalance guarantees at least $|N_A-N_B|$ zero-energy states. This argument explains the zero-energy flat band in regularly inflated RRGs with $L=1,3$, where the graph is bipartite. However, it does not apply to the other cases shown in Fig.~\ref{fig2}, which are generally non-bipartite.

Nevertheless, flat bands are observed in all inflated random graphs, including the non-bipartite cases. As expected, in the absence of inflation, most seed graphs do not exhibit flat bands, with the exception of the scale-free graph, which will be discussed later.

A clear pattern emerges from Fig.~\ref{fig2}. For uniform inflation, flat bands appear with a degeneracy that, for a given seed graph, does not depend on $L$. However, the energies at which these flat bands occur do depend on $L$. Specifically, the flat-band energies are $\varepsilon=0$ for $L=1$, $\varepsilon=\pm 1$ for $L=2$, and $\varepsilon=0,\pm\sqrt{2}$ for $L=3$.

For random inflation, with the exception of the VTG case, the degeneracies vary as functions of both $\bar L$ and energy. Significant flat-band are seen for energies of $\varepsilon=0, \pm 1$ for $\bar L=1$, $\varepsilon=0,\pm 1, \pm\sqrt{2}$ for $\bar L=2$, and $\varepsilon=0,\pm 1,\pm\sqrt{2}, \pm (1+\sqrt{5})/2$ for $\bar L=3$.

The energies at which flat bands appear can be understood from the formation of states localized on the inflated chains and decoupled from the rest of the graph. A finite tight-binding chain of length $L$ has isolated eigenvalues
\begin{equation}
\tilde \varepsilon_m = -2t\cos\!\left(\frac{m\pi}{L+1}\right),
\qquad m=1,\dots,L.
\label{chain}
\end{equation}
If suitable superpositions of these chain eigenmodes satisfy a destructive-interference condition at the original seed sites, they decouple from the rest of the graph and form flat bands. Thus, the flat bands arise from local interference constraints. This explains the appearance of additional flat bands for larger values of $\bar L$, as longer chains support a larger number of internal eigenmodes. It also explains the reduction in the degeneracy of a given flat band in the randomly inflated case: only chains that support the same eigenenergy can participate in the corresponding destructive-interference pattern.

Thus, the general mechanism is local destructive interference at the original seed sites. In an edge-inflated graph, compact localized states (CLS) may be constructed on the inflated chains such that their amplitudes vanish on the junctions connecting them to the rest of the graph \cite{Inui1994,Bergman2008,Flach2014,Leykam2018}. The Schr\"odinger equation at an original site $o\in O$ is
\begin{equation}
\sum_{c\ni o} -t\,\psi^{(c)}_{1} = \varepsilon\,\phi_o,
\label{interference}
\end{equation}
where $\phi_o$ is the amplitude on the seed site $o$, and $\psi^{(c)}_j$ is the amplitude on site $j$ of chain $c$, with $j=1$ denoting the site closest to $o$. For a zero-energy state with vanishing amplitude on the seed sites, this equation reduces to
\begin{equation}
\sum_{c\ni o} -t\,\psi^{(c)}_{1} = 0
\qquad \forall o\in O.
\end{equation}
Thus, the zero eigenstates are obtained by assigning amplitudes on the inflated chains so that the net amplitude entering each original site cancels. The counting of such states is then controlled by the number of independent chain amplitudes relative to the number of independent junction constraints.

The comparison across these distinct graph families therefore suggests that the main spectral features observed in Fig.~\ref{fig2} are governed primarily by the inflation procedure and the resulting filamentary junction geometry, rather than by the detailed topology of the original seed graph. This observation motivates a comparison with the relation between zero-energy states and maximum matching in trees. Specifically, the nullity $\eta(T)$, i.e., the number of zero-energy states of the Hamiltonian associated with the adjacency matrix $A(T)$ of a tree $T$, satisfies $\eta(T)=N-2\nu(T)$, where $\nu(T)$ is the maximum matching number of the tree [see Fig.~\ref{fig1}(e,f)].


\section{Maximum matching and nullity}
\label{s5}

We now take a closer look at the relation between the number of zero-energy states,
\begin{equation}
\eta(G)=\dim\ker A(G).
\end{equation}
Throughout, $\eta(G)$ denotes the number of zero-energy eigenstates of the Hamiltonian associated with $G$. For uniform real hopping, this coincides with the standard graph nullity, $\dim\ker A(G)$. The maximum matching number is $\nu(G)$.  The quantity $N-2\nu(G)$ is the minimum number of vertices that remain unmatched in an optimal hard-dimer covering of the graph.  For a tree or a forest (disconnected trees), the relation between $N-2\nu(G)$ and the nullity is exact.  For graphs containing loops, the situation is more subtle, and it is useful to distinguish the exact statements from the generic properties expected for random graphs.

\subsection{Exact result for trees and forests}

Consider first a tree $T$ with $N$ vertices. Choose a leaf $v$ and its unique neighbor $u$. Remove both vertices $u$ and $v$, together with all edges connected to them. The remaining graph is a smaller forest, which is denoted by $T'$.

At zero energy, the equation at the leaf $v$ is
\begin{equation}
(A(T)\psi)_v=\psi_u=0,
\end{equation}
so every zero mode satisfies
\begin{equation}
\psi_u=0.
\end{equation}
The equation at $u$ then determines the amplitude on the leaf $v$,
\begin{equation}
\psi_v=
-\sum_{\substack{w\sim u\\w\neq v}}\psi_w,
\label{leaf_amp}
\end{equation}
where $w\sim u$ means that $w$ is a nearest neighbor of $u$. 

The important point is that this establishes a one-to-one correspondence between the zero eigenstates of $T$ and those of $T'$.  Indeed, restricting any zero eigenstate of $T$ to the vertices of $T'$ gives a zero eigenstate of $T'$, because every term removed from the remaining equations is proportional to $\psi_u=0$.  Conversely, any zero eigenstate of $T'$ can be uniquely extended to $T$ by setting $\psi_u=0$ and determining $\psi_v$ from Eq.~(\ref{leaf_amp}).  Hence
\begin{equation}
\eta(T)=\eta(T').
\label{nullity_leaf}
\end{equation}

$\nu(T)$ transforms in the same way. There always exists a maximum matching containing the leaf edge $(u,v)$.  To see this, start from any maximum matching.  If $(u,v)$ is already present there is nothing to prove.  If $v$ is unmatched and $u$ is matched to another vertex $w$, the edge $(u,w)$ can be replaced by $(u,v)$ without changing the size of the matching.  For maximum matching, both $u$ and $v$ can not be unmatched, since the edge $(u,v)$ could be added.  Thus
\begin{equation}
\nu(T)=1+\nu(T').
\label{matching_leaf}
\end{equation}
Since $N'=N-2$, Eqs.~(\ref{nullity_leaf}) and (\ref{matching_leaf}) imply
\begin{equation}
N-2\nu(T)
=
N'-2\nu(T').
\end{equation}

Thus both the nullity and the matching deficiency are invariant under the same leaf-pair elimination.  Repeating the procedure until no edges remain produces a set $F_0$ of isolated vertices.  For this final graph,
\begin{equation}
\nu(F_0)=0,
\qquad
\eta(F_0)=|V(F_0)|.
\end{equation}
Since each elimination removes one matched edge and two vertices,
\begin{equation}
|V(F_0)|=N-2\nu(T).
\end{equation}
Consequently,
\begin{equation}
\eta(T)=N-2\nu(T)
\label{tree_matching}
\end{equation}
for every tree, and therefore for every forest \cite{Godsil1981,Godsil1995,Brouwer2012}.

The zero eigenstates of a tree or forest reside on the sites left uncovered by an optimal dimer covering.

\subsection{Exact reduction of an inflated graph}

We now consider an edge-inflated graph $G$ obtained from a seed graph
\begin{equation}
G_o=(V_o,E_o),
\qquad N_o=|V_o|.
\end{equation}
The inflation procedure replaces a seed edge $e=(u,v)$ by the chain
\begin{equation}
u-x^{(e)}_1-x^{(e)}_2-\cdots-x^{(e)}_{\ell_e}-v,
\label{inflated_path}
\end{equation}
where $\ell_e$ is the number of inserted vertices.

Unlike a tree, the seed graph may contain loops, so the leaf-elimination argument cannot in general be applied globally.  Nevertheless, the zero-energy equations at the inserted vertices along each inflated edge can be solved exactly.  Denoting the amplitudes on the original seed vertices by
\begin{equation}
\phi_o\equiv\psi_o,
\qquad o\in V_o.
\end{equation}
On an internal chain site,
\begin{equation}
\psi_{j-1}+\psi_{j+1}=0,
\end{equation}
and hence
\begin{equation}
\psi_{j+1}=-\psi_{j-1}.
\label{zero_recursion}
\end{equation}

It is now useful to distinguish chains with odd and even numbers of inserted sites.

For an odd chain,
\begin{equation}
\ell_e=2s_e+1.
\end{equation}
Choose an orientation for the chain, $u(e)\rightarrow v(e)$, and define
\begin{equation}
a_e\equiv\psi^{(e)}_1.
\end{equation}
Equation~(\ref{zero_recursion}) gives
\begin{align}
\psi^{(e)}_{2r} &=(-1)^r\phi_{u(e)},\\
\psi^{(e)}_{2r+1}&=(-1)^r a_e.
\end{align}
The equation at the last internal site imposes the endpoint condition
\begin{equation}
\phi_{u(e)}+(-1)^{s_e}\phi_{v(e)}=0.
\label{odd_endpoint}
\end{equation}
Thus every odd chain leaves one independent chain amplitude $a_e$, but simultaneously imposes one relation between the amplitudes on its two original endpoints.

Defining the number of odd chains, $M_{\rm odd}$, by
\begin{equation}
E_{\rm odd}
=
\{e\in E_o:\ell_e\ {\rm is\ odd}\},
\qquad
M_{\rm odd}=|E_{\rm odd}|,
\end{equation}
and collect the corresponding amplitudes in
\begin{equation}
a=\{a_e\}_{e\in E_{\rm odd}}.
\end{equation}
Defining the $N_o\times M_{\rm odd}$ matrix $Q$ by
\begin{equation}
Q_{oe}=
\begin{cases}
1, & o=u(e),\\[3pt]
(-1)^{s_e}, & o=v(e),\\[3pt]
0, & o\notin e.
\end{cases}
\label{Qdef}
\end{equation}
Equation~(\ref{odd_endpoint}) for all odd chains is then simply
\begin{equation}
Q^{T}\phi=0.
\label{QTphi}
\end{equation}

For an even chain,
\begin{equation}
\ell_e=2s_e.
\end{equation}
There is no independent residual chain amplitude.  Solving the internal equations instead produces an effective zero-energy coupling between its two original endpoints.  For unit hopping this coupling has sign $(-1)^{s_e}$.  We therefore introduce an $N_o\times N_o$ symmetric matrix $S$, whose off-diagonal matrix element associated with an even inflated edge $e=(u,v)$ is
\begin{equation}
S_{uv}=S_{vu}=(-1)^{s_e}.
\label{Sdef}
\end{equation}

The zero-energy equations at the original vertices now read
\begin{equation}
S\phi+Qa=0.
\label{junction_general}
\end{equation}
Together with Eq.~(\ref{QTphi}), the complete zero-energy problem is therefore reduced to
\begin{equation}
\begin{pmatrix}
S & Q\\
Q^T & 0
\end{pmatrix}
\begin{pmatrix}
\phi\\
a
\end{pmatrix}
=0.
\label{Heff}
\end{equation}
We denote the reduced matrix in Eq.~(\ref{Heff}) by $H_{r}$.

This reduction is exact: every zero mode of the original inflated graph determines a unique solution $(\phi,a)$ of Eq.~(\ref{Heff}), and, conversely, the amplitudes on all eliminated chain sites are uniquely reconstructed from $(\phi,a)$ using Eq.~(\ref{zero_recursion}).  Consequently,
\begin{equation}
\eta(G)=\dim\ker H_{r}.
\label{nullity_reduced}
\end{equation}

This formulation separates the two effects of inflation. Each odd chain generates one residual zero-energy variable $a_e$, while even chains generate effective couplings between the original seed vertices.

\subsection{Matching reduction}

There is an analogous exact simplification of the matching problem.  Replacing an edge $(u,v)$ by
\begin{equation}
u-x_1-x_2-v
\end{equation}
increases the number of vertices by two and the maximum matching number by one.  Hence the matching deficiency, $N-2\nu(G)$, is unchanged by inserting or removing two consecutive internal vertices along an edge.

Applying this operation repeatedly reduces every inflated chain to either one additional vertex or none. An odd number of inserted vertices reduces to a two-edge path $u-y_e-v$ containing one auxiliary vertex $y_e$, while an even number of inserted vertices reduces to a direct edge between the two seed vertices.

The reduced graph, $G_{r}$, contains $N_{r}=N_o+M_{\rm odd}$ vertices and satisfies $N-2\nu(G)=N_{r}-2\nu(G_{r})$. This reduced graph is the starting point for comparing nullity and maximum matching in an inflated graph.

\subsection{All-even inflation}

When all inflated chains contain an even number of inserted sites, i.e., $L=2,4,\ldots$, we have $a=\emptyset$, so the matrix $Q$ is absent from the zero-energy reduction and the reduced Hamiltonian becomes $H_r=S$. Since $S$ has the same connectivity structure as the seed graph $G_o$, apart from the $\pm1$ prefactors associated with individual matrix elements, one expects its nullity to be similar to that of the corresponding uninflated seed graph. As shown in Fig.~\ref{fig2}, the nullity of most of the random seed graphs is close to zero. A notable exception is the SFG, which exhibits a finite nullity even without inflation and will be discussed elsewhere \cite{berkovits2026a}.

A vanishing nullity for all-even inflation does not imply the absence of flat bands at nonzero energies. For uniform even inflation, $L=2,4,\ldots$, the possible flat-band energies are determined by the eigenvalues $\tilde{\varepsilon}_m$ of the corresponding isolated chain, as defined in Eq.~(\ref{chain}). At one of these energies, the amplitudes on the original seed vertices may vanish, while each inflated edge supports the corresponding isolated-chain eigenmode. The resulting eigenstate is therefore specified by one amplitude for each inflated chain, subject to destructive-interference conditions at the original seed vertices.

Equivalently, one may shift the energy by $\tilde{\varepsilon}_m$ and regard the corresponding chain eigenmode as an effective auxiliary degree of freedom at zero energy. In this representation, the problem takes the same form as an $L=1$ inflated graph, with an incidence-like constraint matrix $Q_m$ whose matrix elements are determined by the amplitudes of the $m$th chain eigenmode at its two ends. Denoting the number of even inflated chains by $M_{\rm even}$, the degeneracy of the corresponding flat band is
\begin{equation}
\dim\ker Q_m=M_{\rm even}-\rank Q_m.
\end{equation}
For uniform all-even inflation, $M_{\rm even}=M_o$ is simply the number of edges in the seed graph. For a generic random seed graph, one expects $\rank Q_m=N_o$, giving an extensive flat-band degeneracy $M_{\rm even}-N_o$. Special coherent relations among the junction constraints may reduce the rank by an $O(1)$ amount and thereby generate a small number of additional states.

For example, for $L=2$ the isolated chain has the two eigenvalues $\tilde{\varepsilon}_m=\pm1$. The corresponding eigenmodes have either equal or opposite amplitudes on the two end sites of the chain. Consequently, in the absence of magnetic flux, the associated matrices $Q_m$ have signless- or oriented-incidence forms, respectively. For a connected non-bipartite random seed graph, the signless incidence matrix is generically of full row rank, whereas the oriented incidence matrix has rank $N_o-1$. The two nonzero flat bands therefore have the same extensive degeneracy, although their degeneracies may differ by an $O(1)$ contribution associated with this special rank deficiency.

Random magnetic flux does not remove these nonzero-energy flat bands. Since each inflated chain is an open one-dimensional segment, the Peierls phases along an individual chain can be eliminated by a local gauge transformation and therefore do not change its isolated eigenvalues $\tilde{\varepsilon}_m$. The flux changes the relative complex phases with which the chain eigenmodes enter the junction constraints. For generic random phases, coherent algebraic dependencies are removed and the corresponding matrices $Q_m$ attain full row rank. Random flux can therefore remove additional $O(1)$ states associated with special rank deficiencies, but leaves both the flat-band energies and their extensive degeneracy $M_{\rm even}-N_o$ unchanged.

Thus, for uniform inflation, the isolated-chain spectrum determines the possible flat-band energies, whereas the rank of the corresponding junction-constraint matrix $Q_m$ determines their degeneracies. The same principle will be used below for all-odd inflation.

\subsection{All-odd inflation and structural rank}

A particularly transparent situation occurs when every inflated edge contains an odd number of inserted sites. In this case $S=0$, and the reduced Hamiltonian takes the bipartite form
\begin{equation}
H_r=
\begin{pmatrix}
0 & Q\\
Q^T & 0
\end{pmatrix}.
\label{bipartite_reduced}
\end{equation}
The two sublattices consist of the $N_o$ original vertices and the $M_{\rm odd}=M_o$ residual chain vertices, respectively. Since all inflated chains are odd, $M_{\rm odd}$ is simply the number of edges in the seed graph.

Define $r=\rank Q$. The zero modes of $H_r$ consist of vectors in the kernels of $Q$ and $Q^T$. Therefore
\begin{align}
\eta(G)
&=\dim\ker Q+\dim\ker Q^T \nonumber\\
&=(M_{\rm odd}-r)+(N_o-r),\nonumber\\
&=M_{\rm odd}+N_o-2\rank Q.
\label{nullity_allodd}
\end{align}
This expression may be rewritten as
\begin{equation}
\eta(G) = M_{\rm odd}-N_o + 2\left(N_o-\rank Q\right).
\label{nullity_imbalance}
\end{equation}
Thus the bipartite sublattice imbalance guarantees at least
\begin{equation}
\eta(G)\ge M_{\rm odd}-N_o
\label{bipartite_bound}
\end{equation}
zero eigenvalues. Equality holds when $\rank Q=N_o$.

The maximum matching can be determined directly from the bipartite structure. Assume that the seed graph is connected and that every original vertex has degree of at least two. Choose any subset $\mathcal S$ of original vertices and denote by $\Gamma(\mathcal S)$ the set of odd chains connected to at least one vertex in $\mathcal S$. Defining $E_{\rm int}(\mathcal S)$ as the number of seed edges joining two vertices within $\mathcal S$, and $E_{\rm out}(\mathcal S)$ as the number of edges connecting $\mathcal S$ to the rest of the graph, one has
\begin{equation}
\sum_{v\in\mathcal S} k_v = 2E_{\rm int}(\mathcal S)+E_{\rm out}(\mathcal S) \ge 2|\mathcal S|,
\end{equation}
where the final inequality follows from $k_v\ge2$. On the other hand,
\begin{equation}
|\Gamma(\mathcal S)| = E_{\rm int}(\mathcal S)+E_{\rm out}(\mathcal S).
\end{equation}
Since
\begin{equation}
2|\Gamma(\mathcal S)|
=
2E_{\rm int}(\mathcal S)+2E_{\rm out}(\mathcal S)
\ge
2E_{\rm int}(\mathcal S)+E_{\rm out}(\mathcal S),
\end{equation}
one obtains Hall's condition,
\begin{equation}
|\Gamma(\mathcal S)|\ge|\mathcal S|.
\end{equation}
This condition requires that every subset $\mathcal S$ of original vertices be connected collectively to at least $|\mathcal S|$ distinct chain vertices. When it is satisfied, there exists a matching that covers all $N_o$ original vertices.

The maximum matching of the reduced graph therefore contains $N_o$ edges, leaving
\begin{equation}
M_{\rm odd}-N_o
\end{equation}
vertices unmatched. Since the elimination of pairs of internal chain sites does not change the matching deficiency, the original inflated graph satisfies
\begin{equation}
N-2\nu(G)=M_{\rm odd}-N_o.
\label{matching_allodd}
\end{equation}
Equivalently,
\begin{equation}
\nu(G)=\frac{N+N_o-M_{\rm odd}}{2}.
\label{nu_allodd}
\end{equation}

Combining Eqs.~(\ref{nullity_imbalance}) and (\ref{matching_allodd}) results in
\begin{equation}
\eta(G)-[N-2\nu(G)] = 2\left(N_o-\rank Q\right).
\label{allodd_difference}
\end{equation}
Hence
\begin{equation}
\eta(G)\ge N-2\nu(G),
\end{equation}
with equality if $\rank Q=N_o$.

The same result may be expressed in terms of the structural rank $\operatorname{sprank} Q$, i.e., the maximum rank compatible with the zero-nonzero pattern of $Q$. The zero-nonzero pattern of $Q$ defines the same bipartite incidence graph used above in the matching argument. Since Hall's condition guarantees a matching that covers all $N_o$ row vertices,
\begin{equation}
\operatorname{sprank}Q=N_o.
\end{equation}
Therefore the full-row-rank condition is equivalent to
\begin{equation}
\rank Q=\operatorname{sprank}Q=N_o.
\label{rank_saturation_allodd}
\end{equation}
For a generic random seed graph, one expects the structural rank to be saturated because random connectivity and the signs or phases generated along the different chains suppress coherent algebraic cancellations. In this generic-rank regime,
\begin{equation}
\eta(G)=M_{\rm odd}-N_o=N-2\nu(G).
\label{allodd_identity}
\end{equation}

An instructive exception occurs for uniform $L=3$ inflation in the absence of magnetic flux. In this case the zero-energy eigenstate of an isolated chain has opposite amplitudes on its first and last sites. Consequently, every column of $Q$ contains one $+1$ and one $-1$, so that $Q$ is an oriented incidence matrix of the seed graph. For a connected seed graph,
\begin{equation}
\rank Q=N_o-1,
\end{equation}
and Eq.~(\ref{nullity_imbalance}) gives
\begin{equation}
\eta(G)=M_{\rm odd}-N_o+2.
\label{L3_zero_flux}
\end{equation}
Thus, for $L=3$ and zero flux, two additional zero modes occur beyond the sublattice-imbalance contribution $M_{\rm odd}-N_o$. These states originate from the coherent sign relation associated with the oriented-incidence structure and are not required by the bipartite imbalance itself.

Random magnetic flux removes this special algebraic dependence without breaking the bipartite structure. In the presence of Peierls phases, the transpose $Q^T$ in Eq.~(\ref{bipartite_reduced}) is replaced by the Hermitian conjugate $Q^\dagger$, while the rank counting remains unchanged. The relative complex phases with which a chain couples to its two endpoints destroy the simple $(+1,-1)$ column structure, and for generic random flux one expects
\begin{equation}
\rank Q=N_o.
\end{equation}
The two additional states in Eq.~(\ref{L3_zero_flux}) are therefore removed, leaving
\begin{equation}
\eta(G)=M_{\rm odd}-N_o,
\end{equation}
in agreement with both the bipartite sublattice imbalance and the maximum-matching result in Eq.~(\ref{matching_allodd}). Thus random flux leaves the extensive zero-energy flat band intact while removing the two additional states associated with the oriented-incidence rank deficiency.

Uniform all-odd inflation also produces flat bands at the nonzero eigenvalues $\tilde{\varepsilon}_m$ of the isolated chain. For example, for $L=3$,
\begin{equation}
\tilde{\varepsilon}_m=0,\pm\sqrt{2}.
\end{equation}
The two nonzero chain eigenmodes have equal end-site amplitudes, so that in the absence of magnetic flux their corresponding matrices $Q_m$ have a signless-incidence structure. For a generic connected non-bipartite random seed graph, these matrices have full row rank, and the two nonzero flat bands therefore have degeneracy
\begin{equation}
M_{\rm odd}-N_o.
\end{equation}
The zero-energy flat band differs by the two additional states in Eq.~(\ref{L3_zero_flux}) only in the zero-flux case.

As discussed above, random magnetic flux does not change the isolated-chain eigenvalues $\tilde{\varepsilon}_m$, but it generically removes special algebraic rank deficiencies in the corresponding junction constraints. For $L=3$, the nonzero-energy matrices $Q_m$ are already of full row rank in the generic zero-flux case, so their flat-band degeneracies remain $M_{\rm odd}-N_o$. The zero-energy constraint matrix instead changes from rank $N_o-1$ to rank $N_o$, reducing its degeneracy by two. Consequently, in the presence of generic random flux, the flat bands at $\tilde{\varepsilon}_m=0,\pm\sqrt{2}$ all have the same degeneracy $M_{\rm odd}-N_o$.

The same reasoning applies to any uniformly inflated all-odd graph: the isolated-chain eigenvalues determine the possible flat-band energies, while the ranks of the corresponding junction-constraint matrices determine their degeneracies.

\subsection{Mixed parity and random seed graphs}

When both even and odd inflated chains are present, $S\neq0$ in general and the reduced graph is no longer necessarily bipartite. The exact nullity is then given by Eq.~(\ref{nullity_reduced}),
\begin{equation}
\eta(G)= \dim\ker
\begin{pmatrix}
S&Q\\
Q^T&0
\end{pmatrix},
\end{equation}
together with the matching relation $N-2\nu(G)=N_{r}-2\nu(G_{r})$. In this more general case, the matching number cannot be obtained from $\operatorname{sprank}Q$ alone, because the even chains generate direct edges between the original vertices. In the presence of Peierls phases, the reduced matrix is Hermitian and $Q^T$ is replaced by $Q^\dagger$.

This distinction is important: once both chain parities are present and the reduced graph is generally non-bipartite, there is no universal identity between adjacency-matrix nullity and matching deficiency. The two quantities can differ because closed loops generate algebraic constraints that have no analogue in the tree proof.

For the random seed graphs considered here, however, one expects coherent algebraic dependencies to be strongly suppressed. Random connectivity, irregular chain lengths, and varying signs or phases suppress the systematic cancellations characteristic of ordered lattices. Results on the rank and invertibility of sparse random-graph adjacency matrices support the general expectation that such matrices are generically of high rank \cite{Bordenave2011,Huang2021}. These results do not by themselves constitute a proof of
\begin{equation}
\eta(G)=N-2\nu(G)
\end{equation}
for the present inflated ensembles; rather, they motivate the generic-rank hypothesis underlying the numerical comparison below.

We therefore distinguish three situations. For forests, Eq.~(\ref{tree_matching}) is an exact theorem. For all-odd inflated graphs, equality follows when the corresponding constraint matrix attains full row rank. For general randomly inflated graphs containing both chain parities, the close correspondence between $\eta(G)$ and $N-2\nu(G)$ is a genericity conjecture to be tested numerically rather than a consequence of the tree identity alone.

This viewpoint also explains why ordered inflated lattices can behave differently \cite{Berkovits2026}. Symmetry, periodicity, repeated motifs, or coherent loop phases may produce exact linear dependencies among the reduced zero-energy constraints. Such dependencies generate zero eigenstates beyond those expected from a purely combinatorial matching count.


\begin{figure}
    \includegraphics[width=0.5\textwidth]{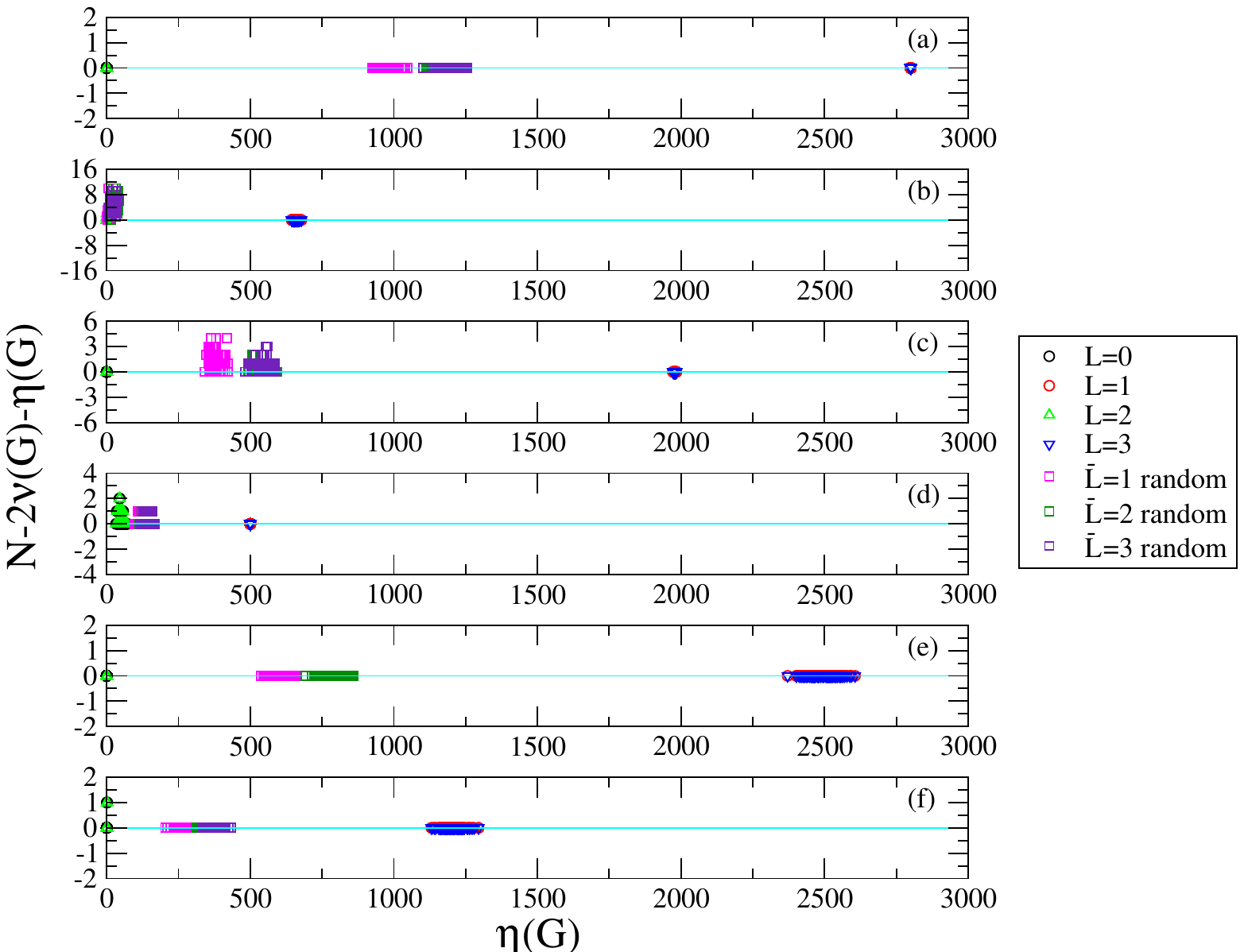}
    \caption{Correspondence between the nullity of a graph, $\eta(G)$, and its maximum matching, $\nu(G)$, for $100$ independent realizations of each network class, using the same parameter values as in Fig.~\ref{fig2}. For each realization, we evaluate the deviation from the conjectured relation $\eta(G)=N-2\nu(G)$, quantified by $N-2\nu(G)-\eta(G)$, for different values of the uniform inflation length $L$ and, in the random case, the average inflation length $\bar L$. The panels correspond to (a) RRG, (b) VTG, (c) DTG, (d) SFG, (e) SWG, and (f) ERG. For the RRG, the relation $\eta(G)=N-2\nu(G)$ holds exactly for all values of $L$ and $\bar L$. The ERG shows a very similar pattern, except for $L=0,2$, where no zero eigenvalues are present but the matching deficiency may overestimate the nullity by one. For the SWG, the relation holds exactly except for some realizations with random inflation, $\bar L=1,3$. For the planar random graphs, the relation holds exactly for odd $L$ and up deviations on order one for the even $L$ in the VTG, while deviations of $O(10)$ appear for the random inflation. In the DTG, the relation holds exactly for all uniformly inflated graphs, but deviations of up to $4$ occur for random inflation. Finally, for the SFG, the relation holds very well in all cases except for $\bar L=0,2$, where deviations of up to two may appear.}
    \label{fig3}
\end{figure}

We now turn to a numerical verification of our conjecture, namely $\eta(G)=N-2\nu(G)$. In all cases, random phases are introduced in order to remove the additional coherent zero-energy states discussed above. Strictly speaking, in this case we calculate the nullity of the Hamiltonian of the graph not of the graph, but as previously discuss, the difference is of $O(1)$ and to avoid introducing cumbersome new notation we will continue to refer to $\eta(G)$. As shown in Fig.~\ref{fig3}, this relation holds very well for all seed and inflated graphs, with deviations of at most $O(1)$, except for the VTG, where deviations of $O(10)$ occasionally occur. Thus, the numerical results strongly support the conjecture, up to small finite-size deviations.

Moreover, in all cases we find $N-2\nu(G)-\eta(G)\ge 0$. For the all-odd cases ($L=1,3$), this behavior is consistent with the expected rank saturation $\operatorname{sprank}Q=\rank Q$ for random seed graphs. By contrast, for randomly inflated graphs with mixed chain parity, this simple equality need not hold, and individual realizations may deviate from $N-2\nu(G)-\eta(G)=0$. The largest deviations from the conjecture occur for the randomly inflated planar graphs ($\bar L$), which are structurally closest to randomly inflated lattice seeds. Similar, and generally larger, deviations were observed for such lattices; see Fig.~13 of Ref.~\onlinecite{Berkovits2026}.

\begin{figure}[t]
\includegraphics[width=0.5\textwidth]{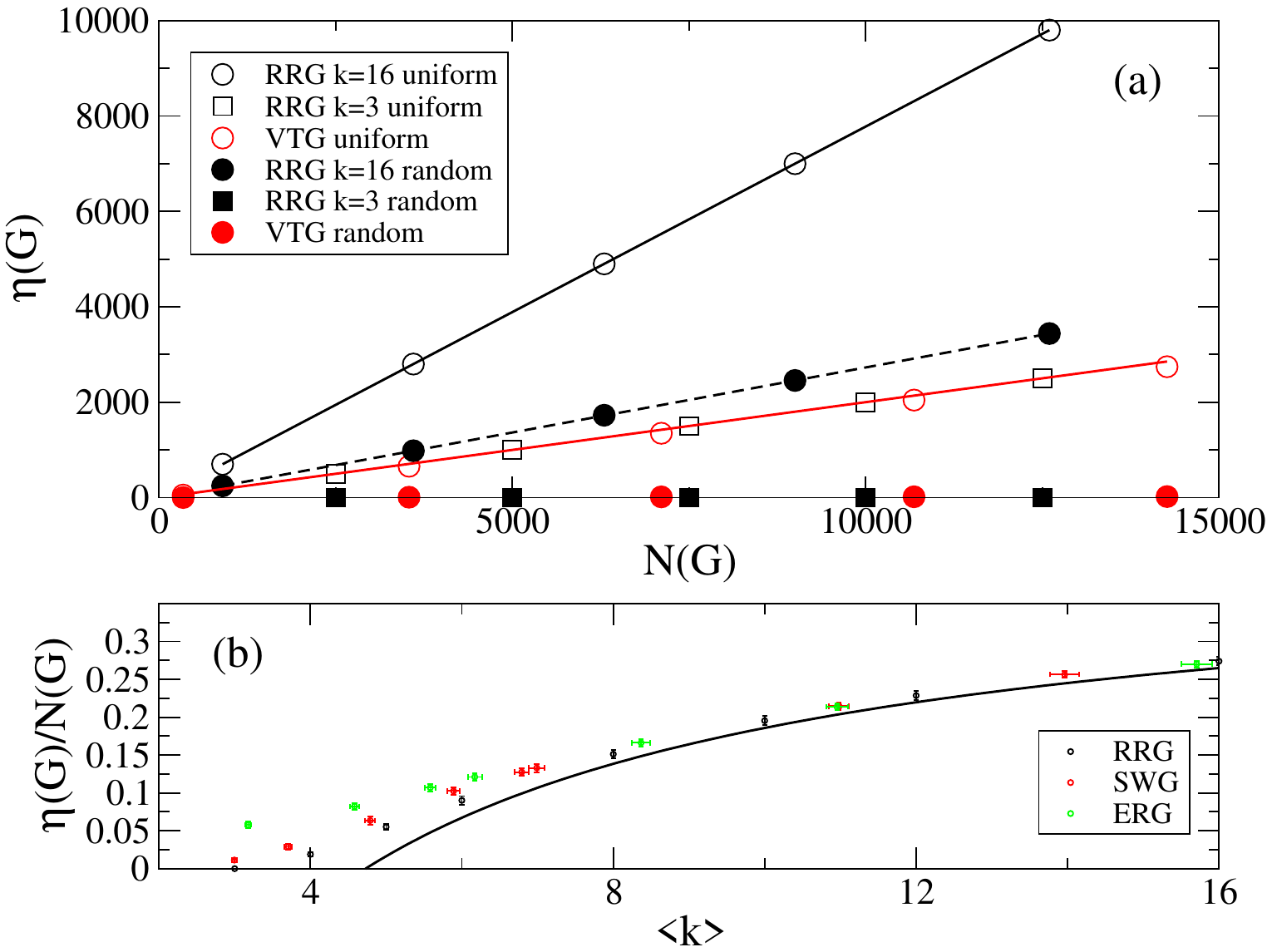}
\caption{(a) The number of zero-energy states $\eta(G)$ as a function of the total number of sites $N(G)$ for RRGs with $k=16$ and $k=3$, and for the VTG, for which $\langle k \rangle \sim 3$. Results are shown for both uniform inflation with $L=1$ and random inflation with $\bar L=1$. The results are averaged over $100$ independent realizations of the seed graph $G_o$. Eq.~(\ref{g_to_n}) for uniform inflation is represented by the solid black ($k=16$) and red ($k=3$) lines. For random inflation, Eq.~(\ref{g_to_n_random}) is represented by the dashed black line for $k=16$, whereas for $k \sim 3$ it predicts a vanishing extensive contribution, as discussed in the text. (b) The ratio $\eta(G)/N(G)$ as a function of the average degree $\langle k\rangle$ of the seed graph for RRG, SWG, and ERG ensembles with random inflation. The curve represents the prediction of Eq.~(\ref{g_to_n_random}).
}
\label{fig3a}
\end{figure}

For uniform inflation, the thermodynamic limit fraction of states belonging to the flat bands can be obtained straightforwardly. As discussed above, for a given inflation length $L$, there are $L$ flat bands and a total of $N=N_o+L M_o$ states, while each flat band contains $M_o-N_o$ degenerate states. Since $M_o=\langle k\rangle N_o/2$, the ratio between the total number of states in the flat bands and the total number of states, denoted by $R_{\rm FB}$, is
\begin{equation}
R_{\rm FB} = \frac{L(M_o-N_o)}{N} = \frac{\langle k\rangle-2}{\langle k\rangle+2/L}.
\label{rfb}
\end{equation}
Thus, for the $\varepsilon=0$ flat band, which is the only flat band present for uniform $L=1$ inflation,
\begin{equation}
  \frac{\eta(G)}{N} = \frac{\langle k\rangle-2}{\langle k\rangle+2}.
  \label{g_to_n}
\end{equation}
This prediction is tested in Fig.~\ref{fig3a}(a) for uniformly inflated ($L=1$) RRG and VTG models of different sizes $N$. Eq.~(\ref{g_to_n}) describes the numerical results very well.

The significantly different values of $\eta(G)$ observed in Fig.~\ref{fig3a}(a) for the randomly inflated graph ensembles can be understood, to leading order, from the number of odd chains relative to the number of original vertices. Defining $P_{\rm odd}$ as the probability that an original edge is inflated by an odd number of sites, the number of odd chains is
\begin{equation}
  M_{\rm odd}\simeq P_{\rm odd} M_o = \frac{P_{\rm odd}\langle k\rangle}{2}N_o .
  \label{modd}
\end{equation}
The matrix $Q$, which couples the $M_{\rm odd}$ odd-chain amplitudes to the $N_o$ original vertices, is expected to have the generic rank $\rank Q \simeq \min(N_o,M_{\rm odd})$. Thus, the contribution of chain-supported zero-energy states may be approximated by $\dim \ker Q \simeq \max(M_{\rm odd}-N_o,0)$, implying that an extensive nullity is expected only when $P_{\rm odd} \langle k \rangle > 2$. Specifically, for $\bar L = 1$,
\begin{equation}
  \frac{\eta(G)}{N} = \frac{\max(P_{\rm odd}\langle k\rangle-2,0)}{\langle k\rangle+2}.
  \label{g_to_n_random}
\end{equation}

This criterion naturally distinguishes between the different behaviors of the randomly inflated graphs shown in Fig.~\ref{fig3a}(a). For an RRG with a high degree, such as $k=16$, one has $M_{\rm odd}\gg N_o$, resulting in an extensive zero-energy degeneracy. By contrast, for an RRG with $k=3$, as well as for the VTG with $\langle k\rangle\simeq3$, one has $M_{\rm odd}<N_o$ for the random-inflation parameters considered here. For $\bar L=1$, $P_{\rm odd}\sim(1-\exp(-2))/2$, while for increasing $\bar L$, $P_{\rm odd}\rightarrow1/2$. Consequently, both the randomly inflated RRG with $k=3$ and the randomly inflated VTG exhibit a negligible number of zero-energy states, in agreement with the numerical results.

A comparison between the prediction of Eq.~(\ref{g_to_n_random}) and the numerical results for the RRG, SWG, and ERG is shown in Fig.~\ref{fig3a}(b). The SFG is discussed separately below. For large values of $\langle k\rangle$, the numerical results for the different graph ensembles follow the prediction of Eq.~(\ref{g_to_n_random}) quite closely. At smaller values of $\langle k\rangle$, deviations from the analytical prediction become more pronounced, and differences between the various graph ensembles emerge. The RRG, for which the degree is fixed, follows Eq.~(\ref{g_to_n_random}) most closely. For the SWG and ERG, where the vertex degrees fluctuate, the behavior depends not only on the average degree but also on its distribution. A more complete description of these cases will therefore require taking the degree fluctuations into account.

Thus, the average degree of the seed graph is the main factor controlling the fraction of degenerate zero-energy states in randomly inflated graphs, while fluctuations in the degree distribution provide additional corrections. Within this picture, increasing the average degree of the seed graph increases the fraction of states belonging to the flat band.

It is interesting to note that for SFGs, which are characterized by a power-law degree distribution, a macroscopic degeneracy is present even in the absence of inflation. Nevertheless, the conjecture remains a reliable predictor of the nullity at $L=0$. Since no comparable degeneracy is observed for ERGs, despite their fluctuating node degrees, this behavior appears to originate from the broad power-law degree distribution characteristic of the SFGs. A more detailed investigation of this issue will be presented elsewhere \cite{berkovits2026a}.

In most practical settings, computing a maximum matching is substantially less expensive than computing the full Hamiltonian spectrum. Calculating the nullity from matching typically scales as $O(M\sqrt{N})$, whereas calculating $\eta(G)$ by full diagonalization has an estimated cost of $O(N^3)$.

\section{Robustness to disorder}
\label{s6}

\begin{figure}[t]
\includegraphics[width=0.5\textwidth]{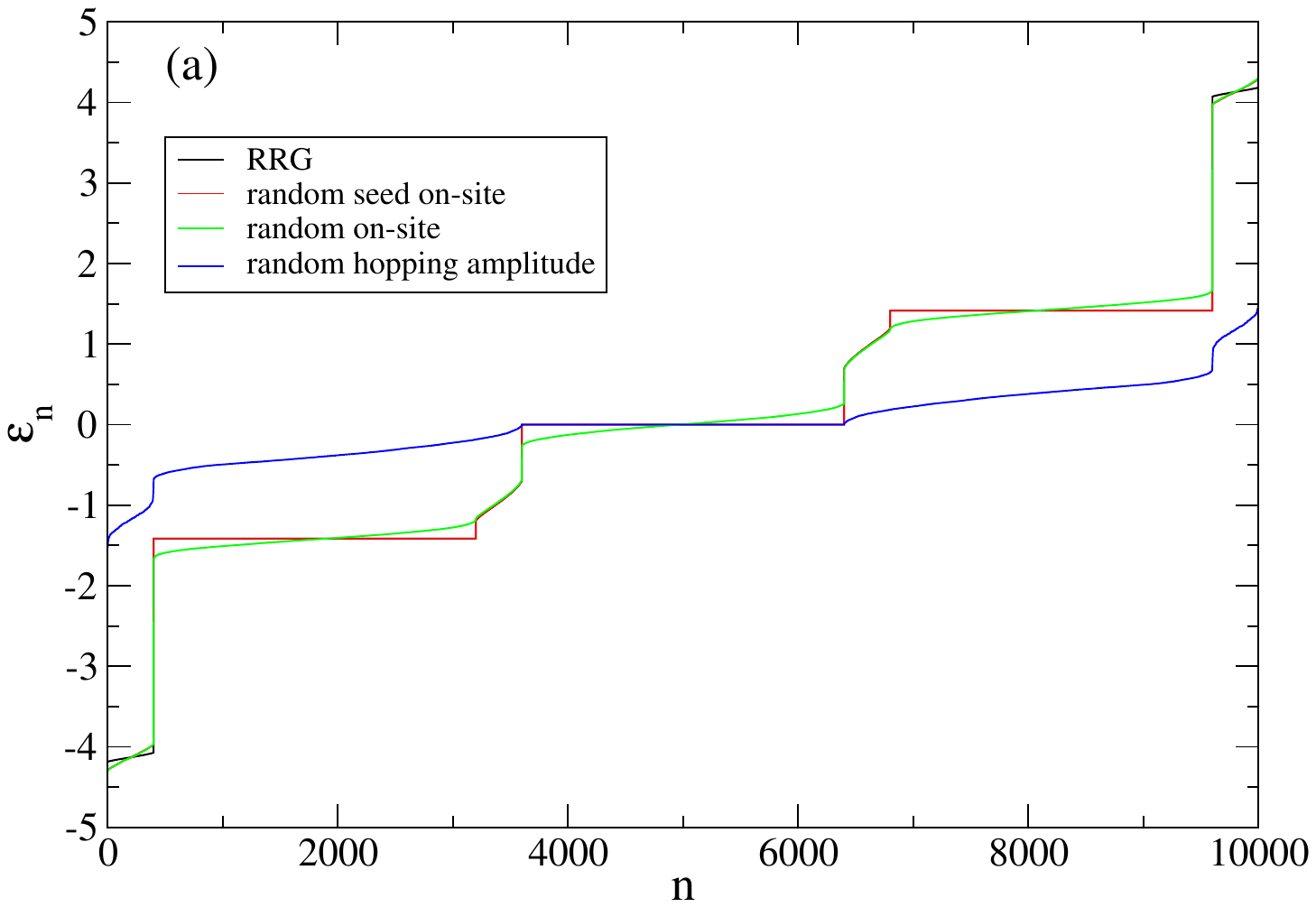}
\includegraphics[width=0.5\textwidth]{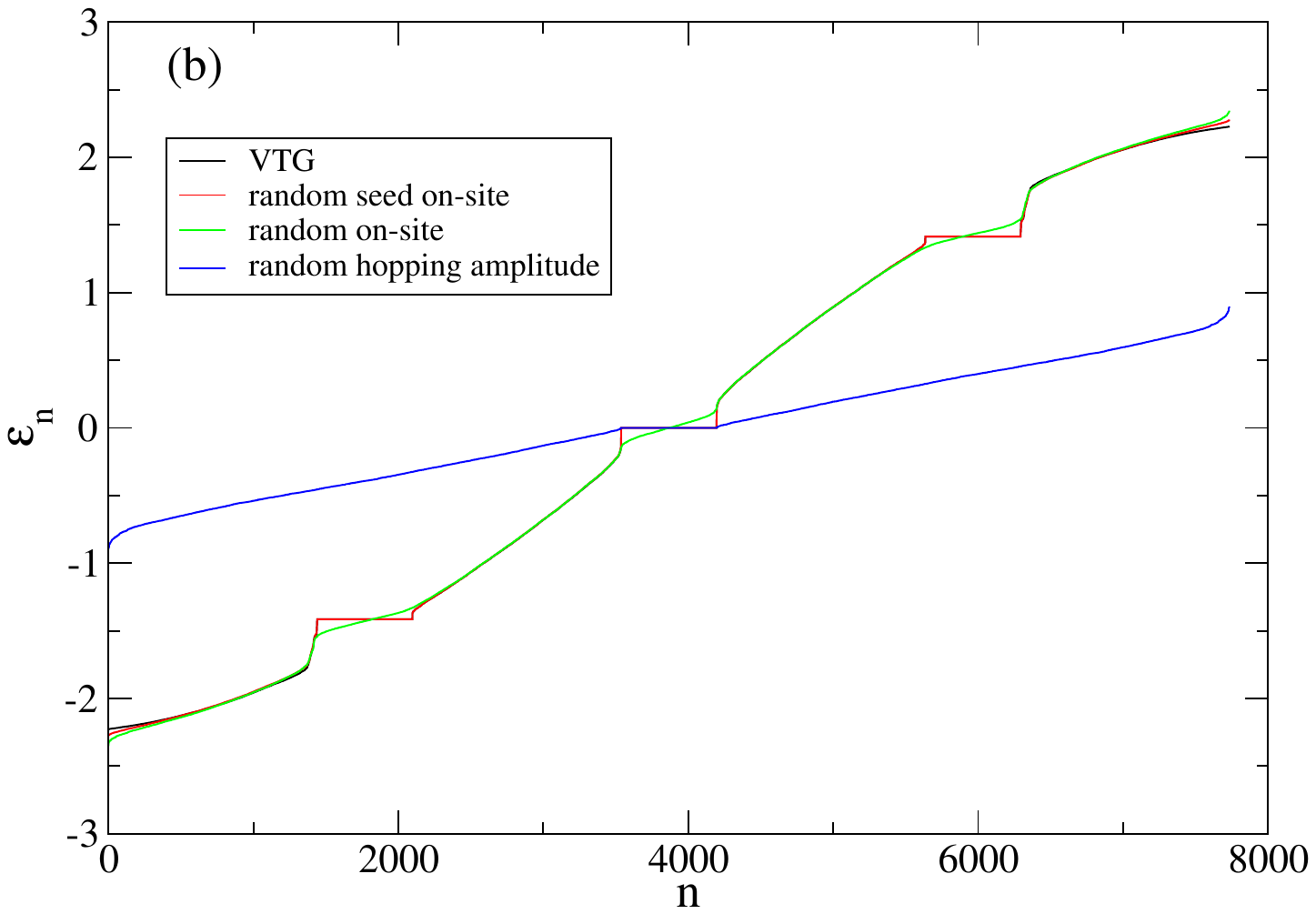}
\caption{The $n$-th energy level $\varepsilon_n$ as a function of the level index $n$ for three types of disorder in (a) RRG and (b) VTG graphs, using the same seed graphs as in Fig.~\ref{fig2}. Black curves correspond to the case with $L=3$ and no disorder. Red curves depict the case in which the on-site energies on the seed sites are chosen from a uniform distribution in the interval $[-0.3,0.3]$, together with random hopping phases corresponding to random magnetic flux. Green curves show the case in which all on-site energies are chosen from the same interval, again with random hopping phases. Blue curves represent the case of random hopping amplitudes chosen from the interval $[-0.5,0.5]$, also with random hopping phases.
}
\label{fig9}
\end{figure}

The robustness of flat bands in inflated random graphs against various types of disorder, including bond disorder, site disorder, and random magnetic flux, is consistent with the behavior observed for inflated lattices \cite{Berkovits2026}. Specifically, as shown in Fig.~\ref{fig9}, most flat bands are broadened by bond disorder, whereas the zero-energy flat band remains protected. Site disorder on the seed sites has no effect on any of the flat bands, while disorder on the chain sites broadens all flat bands. Random magnetic flux leaves the flat-band energies and their leading extensive degeneracies unchanged, although it may remove $O(1)$ states associated with special coherent rank deficiencies. 

\section{Motifs}
\label{s7}

\begin{figure*}[t]
\centering
\includegraphics[width=0.95\textwidth]{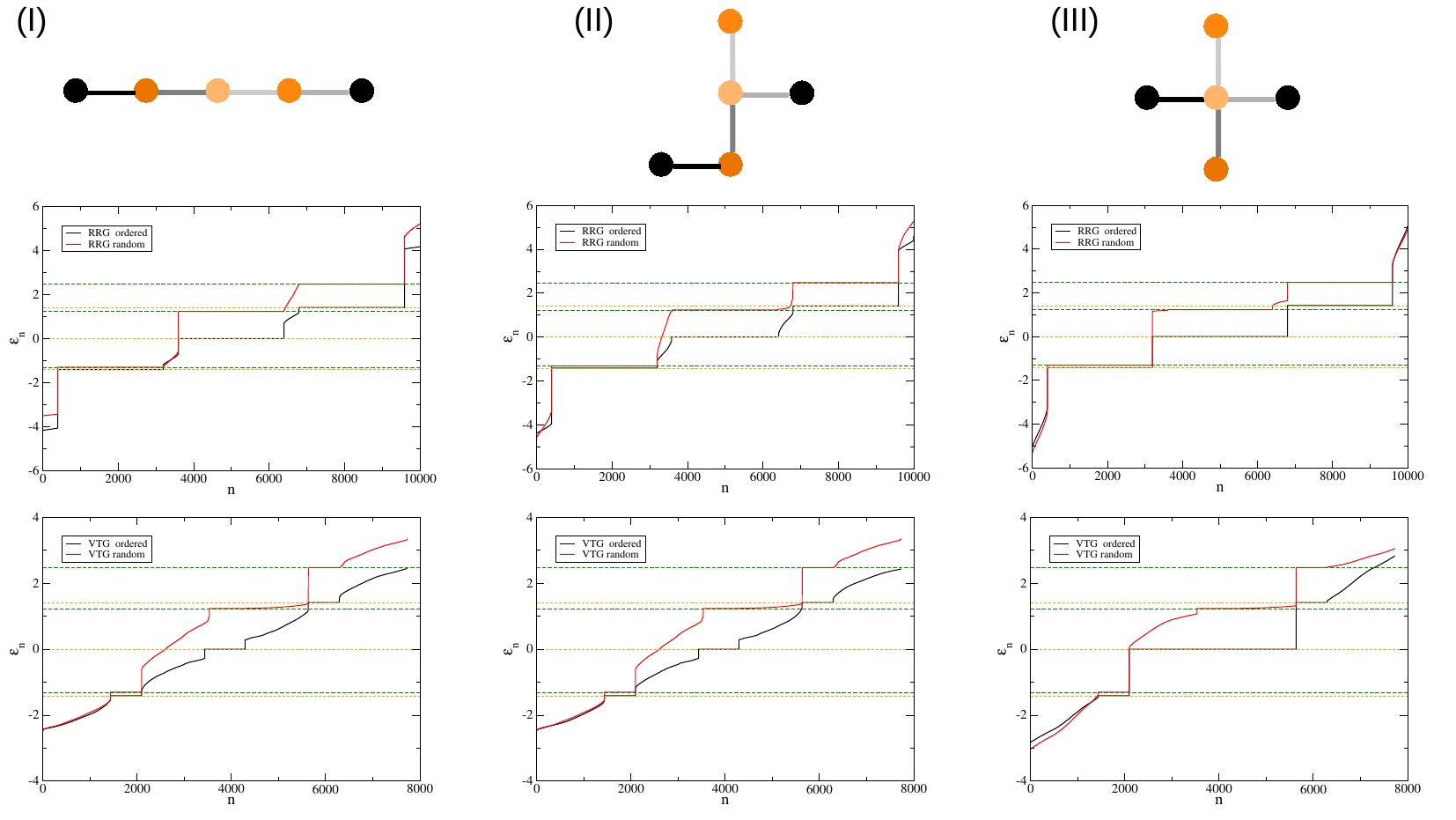}
\caption{The $n$-th energy level $\varepsilon_n$ as a function of the level index $n$ for three different motifs (I, II, and III) in RRG and VTG graphs, using the same seed graphs as in Fig.~\ref{fig2}. These motifs replace the bonds of the seed graphs, whose original sites are shown as black circles, by chains consisting of three nodes (orange shaded circles) connected by bonds (gray shaded lines). Each motif is shown in two versions. In the ordered case, all hopping amplitudes are $-1$ and all on-site energies are zero. In the random case, the different shades of orange indicate different on-site energies, specifically $1.8$, $-0.6$, and $1.2$, while the two hopping amplitudes along the motif are $1.41$ and $-0.41$. The spectra obtained by replacing all edges of a single RRG or VTG seed graph by the ordered motifs (black curves) or random motifs (red curves) are shown in the six panels. The eigenvalues of the corresponding freestanding motifs are indicated by green dashed lines for the ordered case and orange dashed lines for the random case. In all cases, these eigenvalues coincide with one of the flat bands of the inflated graphs.
}
\label{fig5}
\end{figure*}

The inflation procedure considered here can be generalized by replacing the seed bonds with motifs, as shown in Fig.~\ref{fig5}. As long as the motifs contain no closed loops, flat bands appear at energies corresponding to the eigenvalues of the freestanding motifs, similar to the behavior found for chains. As shown in Fig.~\ref{fig5}, the appearance of flat bands at the eigenvalues of the freestanding motifs is insensitive to whether the motif is ordered, with fixed on-site energies and uniform hopping amplitudes, or disordered, with randomly chosen on-site energies and hopping amplitudes, provided that all motifs are identical.

Moreover, for a given realization of a random graph, the degeneracies of the nonzero flat bands are equal. An exception may occur at zero energy, where particular motifs, such as the one shown in Fig.~\ref{fig5}(III), can produce a much larger flat-band degeneracy. Once disorder is introduced within the motif, however, the degeneracy of the zero-energy flat band becomes identical to that of the other flat bands.

\section{Discussion and Conclusions}
\label{s8}

We have shown that edge inflation provides a simple and general mechanism for generating flat bands in random tight-binding networks. By replacing the edges of a random seed graph with finite one-dimensional chains or other motifs, one produces compact localized states whose amplitudes cancel at the original seed vertices. This mechanism does not rely on crystalline order and therefore applies to a broad class of random graph ensembles, including RRGs, VTGs, DTGs, SFGs, SWGs, and ERGs.

The central physical point is that flat bands in these systems are controlled by local constraint structure rather than by global spatial symmetry. The possible flat-band energies are inherited from the local objects replacing the seed bonds. For ordinary edge inflation, these local objects are finite chains, so the flat-band energies are determined by the isolated-chain spectrum. More generally, the same principle applies when the seed bonds are replaced by identical loopless motifs: the eigenvalues of the freestanding motif appear as flat-band energies of the full inflated random graph. Thus, the construction separates the problem into a local spectral part, which fixes the possible flat-band energies, and a global graph-connectivity part, which fixes the number of independent compact localized states.

For the flat-band states, this counting is closely related to the maximum-matching structure of the inflated graph. The numerical results show that, for the random graph ensembles studied here, the nullity $\eta(G)$ is very well described by the matching deficiency $N-2\nu(G)$. This correspondence can be understood by shifting the energy to the eigenvalues of the isolated chains or local motifs and then applying the arguments developed in Sec.~\ref{s5}. 

The fraction of states belonging to the flat bands is controlled primarily by the connectivity of the seed graph. For uniform inflation, the flat-band fraction can be expressed directly in terms of the average degree $\langle k\rangle$, and increases monotonically with increasing connectivity. A similar trend is found for random inflation: the zero-energy degeneracy is governed, to leading order, by the number of odd inflated chains relative to the number of original vertices, implying that an extensive nullity appears only when $P_{\rm odd}\langle k\rangle>2$. The numerical results for the RRG, SWG, and ERG ensembles follow this prediction well at large $\langle k\rangle$, while deviations become more pronounced at lower connectivity, where fluctuations in the degree distribution and other structural correlations become important. Thus, the average degree of the seed graph provides the dominant control parameter for the flat-band fraction, with the detailed degree distribution producing secondary corrections.

The motif construction shows that edge-inflated random graphs are only one member of a broader class of graph-rewritten tight-binding networks. As long as all seed bonds are replaced by identical two-terminal loopless motifs, the flat-band mechanism remains intact. The detailed values of the on-site energies and hopping amplitudes inside the motif affect the flat-band energies, but not the basic existence of the flat bands. In this sense, the flat bands are not tied to fine tuning of a particular chain Hamiltonian, but to the repeated insertion of identical local structures whose amplitudes can be combined to cancel at the seed vertices.

The disorder response further clarifies the nature of these states. Site disorder on the original seed vertices does not affect the flat bands, because the compact localized states have vanishing amplitude on those sites. Random magnetic flux also leaves the flat bands essentially unchanged. By contrast, bond disorder and site disorder on the chain or motif sites generally broaden the flat bands, since they directly perturb the local eigenmodes that form the compact localized states. The zero-energy flat band is more robust against bond disorder, reflecting its special chiral or matching-related origin.

These results suggest that flat bands in random networks should be viewed as generic consequences of local graph rewriting and destructive-interference constraints. They connect compact localized states, graph matching, structural rank, and sparse random-matrix structure within a single framework. The broader implication is that flat-band physics does not require periodicity: it can be engineered in random networks by controlling the local building blocks placed on the edges.

Several open questions remain. First, the role of spatial correlations in planar graphs, such as VTGs and DTGs, should be clarified, since these systems exhibit the largest deviations from the matching-based prediction. Second, extending the analysis from loopless motifs to motifs containing internal loops may reveal additional mechanisms for degeneracy enhancement or flat-band splitting. Third, replacing the local motifs with topological units, such as SSH segments or more complex topological networks, may provide a route toward random-network platforms that host robust localized modes together with nontrivial topological structure. Finally, the effects of electron-electron interactions, particularly when the chemical potential is tuned to a flat band, are expected to give rise to rich many-body behavior.

\begin{acknowledgments}
  I thank S. Flach, A. Andreanov, Y. Kim, R. Cohen, S. Haber, and S. Havlin for insightful discussions.
\end{acknowledgments}

\bibliography{rrg}

\end{document}